\documentclass[prl,twocolumn,english,
superscriptaddress,
floatfix,longbibliography]{revtex4-2}

\pdfoutput=1
\usepackage[utf8]{inputenc}
\usepackage[T1]{fontenc}
\usepackage{braket,xcolor}
\usepackage{graphicx,enumerate,verbatim,bbold}
\usepackage{amsmath,amssymb,amsthm,float,mathrsfs}
\usepackage{dsfont}
\usepackage[normalem]{ulem}
\usepackage{lmodern}
\usepackage{hyperref}
\hypersetup{colorlinks,citecolor=blue,linkcolor=blue,urlcolor=blue}
\usepackage{braket}
\usepackage{multirow}
\usepackage{bbm}
\usepackage{physics}
\usepackage{bm}
\usepackage{tikz}
\usetikzlibrary{patterns}
\usetikzlibrary{fadings}
\usetikzlibrary{quantikz2}
\usepackage[normalem]{ulem}
\usepackage[caption=false]{subfig}
\usepackage{dcolumn}

\newcommand{\circleX}{\textcolor{blue}X}
\newcommand{\circleY}{\textcolor{red}Y}
\newcommand{\circleZ}{\textcolor{black}Z}
\newcommand{\circleI}{\textcolor{gray}\um}

\usepackage[table]{xcolor}

\allowdisplaybreaks

\newcommand{\bqa}{\begin{eqnarray}}
\newcommand{\eqa}{\end{eqnarray}}

\newcommand{\be}{\begin{equation}}
\newcommand{\ee}{\end{equation}}

\usepackage{yhmath} 

\newcommand{\Htarget}{\mathcal{K}_\mathbf{T}}
\newcommand{\Hinit}{\mathcal{K}(0)}
\newcommand{\Hevo}{\mathcal{K}(t)}
\newcommand{\Hevofinal}{\mathcal{K}(T)}

\newcommand{\del}[1]{\hil{{\bf XXX}}}

\newcommand{\hil}[1]{{\color[rgb]{0.6,0,0}{#1}}}
\newcommand{\um}{\mathbbm{1}}

\newcommand{\stb}[3]{%
  \mathbf{#1}\substack{#2\\#3}%
}
\newcommand{\HOBC}{\mathcal{K}^\alpha_{n-\alpha}}
\newcommand{\HPBC}{\mathcal{K}^\alpha_{n}}
\newcommand{\HPBCxy}{\mathsf{K}^\alpha_{n}}
\newcommand{\stab}[3]{\mathbf{#1}_{#2}^{#3}}
\newcommand{\stabb}[4]{(\mathbf{#1}\mathbf{#2})_{#3}^{#4}}

\usepackage{subfig}

\definecolor{nqdcolor}{rgb}{0.5586, 0.0586, 0.4219}

\usepackage{comment}

\newcommand{\Imperial}{Blackett Laboratory, Imperial College London, SW7 2AZ, United Kingdom}

\newcommand{\pku}{State Key Laboratory of Artificial Microstructure and Mesoscopic
Physics, School of Physics, Peking University, 100871 Beijing, China}

\begin{document}
\title{
Quantum simulation of circular cluster interactions in a linear spin chain
}

\author{Alexander van Lomwel}\affiliation{\Imperial}
\author{Hongzheng Zhao}\affiliation{\pku}
\author{Florian Mintert}\affiliation{\Imperial}

\begin{abstract}
The preparation of ground states of Hamiltonians with symmetries that are fundamentally different than those of the underlying device is a central challenge for quantum technologies.
Here, we address this challenge with an analog protocol for the iconic generalized cluster Hamiltonian with the translational symmetry resultant from periodic boundary conditions based on a magnetization-preserving interaction on a linear geometry with open boundary conditions.
Our results show that this goal can be achieved in a control duration that scales only moderately with the number of spins and Hamiltonian interaction complexity.
\end{abstract} 
\maketitle


\paragraph*{Introduction.}
The realization of eigenstates of parent Hamiltonians with sophisticated interactions is crucial for a wide range of applications, including quantum error correction~\cite{kitaev2003fault,bombin2006topological,bravyi2010tradeoffs}, quantum simulation~\cite{georgescu2014quantum,altman2021quantum,wecker2015progress,malley2016scalable,farrell2024scalable} and quantum optimization algorithms~\cite{farhi2014quantum,albash2018adiabatic,zhou2020quantum,nguyen2023quantum}.
Of particular interest are the eigenstates of parent Hamiltonians with periodic boundary conditions (PBC) which suppress edge effects, restore translational symmetry, and that naturally arise in topological error correcting codes such as the toric code~\cite{bombin2012strong,valenti2019hamiltonian}, while allowing bulk behavior and finite-size trends to be studied without additional contributions from physical boundaries~\cite{periwal2021programmable,campostrini2014finite}.

An iconic example is the one-dimensional cluster state defined in terms of a parent Hamiltonian with a three-body interaction that provides a foundational resource for measurement-based quantum computation~\cite{raussendorf2001one}.
Together with its generalizations to higher-order cluster interactions,
this cluster state defines a paradigmatic family of symmetry-protected topological phases, stabilizer resource states, and cluster-Ising critical points~\cite{son2011quantum,smacchia2011statistical,seifnashri2024cluster,lahtinen2015realizing,ohta2016topological,verresen2017one,else2012symmetry}.
With open boundary conditions (OBC), these models support symmetry-protected edge degrees of freedom and an associated boundary-induced ground state degeneracy; PBC remove the physical edges and isolate the translationally invariant bulk state, whose string order, entanglement structure, and phase transitions provide standard diagnostics of the cluster phase~\cite{ohta2016topological,verresen2017one,son2011topological,smacchia2011statistical,perez2008string}.

The core difficulty in creating such cluster states stems from the fact that the three-body or many-body interactions used to define these states are non-existing native interactions in any quantum hardware.
The resort to defining a parent Hamiltonian can be avoided with digital state preparations~\cite{zhang2022digital,gong2019genuine,bluvstein2022quantum,kuznetsova2012yelin,tan2023realizing,iqbal2024topological},
but the limited accuracy of entangling gates in existing devices has resulted in rapidly growing interest in analog approaches~\cite{preskill2018quantum,georgescu2014quantum,altman2021quantum,daley2022practical,flannigan2022propagation}.

The combination of single-qubit driving and two-body interactions can lead to effective three-body or many-body interactions as most notably developed in Floquet engineering~\cite{decker2020floquet,petiziol2021quantum,petiziol2024nonperturbative,koyluoglu2025floquet,wu2026engineering}.
While suitable driving can be designed numerically exactly for sufficiently small systems~\cite{maskara2025programmable,greenaway2024variational},
the construction of effective Hamiltonians is typically based on a perturbative approximation with two distinct aspects that impose severe limitations.
Any attempt to strengthen the magnitude of desired effective processes nearly unavoidably also enhances undesired higher-order effective processes.
The weakness of the driving required for the validity of the perturbative construction necessarily implies a strong signature of the underlying system interactions in the final effective Hamiltonian~\cite{petiziol2021quantum,petiziol2024nonperturbative,eckardt2015high}.
As such, if the underlying system contains a symmetry absent from the desired effective Hamiltonian, weak driving will only show deviations from this symmetry on sufficiently long time scales~\cite{petiziol2024nonperturbative,baumgartner2025hilbert}.

The realization of eigenstates of PBC systems is inherently hardware dependent and introduces an additional challenge of creating symmetry, {\it i.e.} translational, as opposed to breaking symmetry. 
While recent experiments have accessed topological states and periodic geometries using digitally compiled circuits, coherent atom transport, or programmable trapped-ion gates~\cite{satzinger2021realizing,bluvstein2022quantum,haghshenas2026digital},
the ability to modify an interaction geometry by moving qubits or to realize interactions between distant qubits is an exception of selected platforms that is currently out of reach for many platforms, such as registers of superconducting qubits.

In this Letter,
we resolve these challenges by designing analog protocols for the preparation of ground states of generalized cluster Hamiltonians with PBC, based on a linear nearest-neighbor interaction geometry system with OBC, using single-qubit driving to overcome symmetry restrictions
imposed by the underlying interactions.

\paragraph*{Linear spin system.} 

While the numerically exact design of driving schemes for quantum many-body systems is generally an unsurmountable challenge on its own,
it is feasible when the operators that comprise the time-dependent system Hamiltonian $H(t)$  generate a small Lie algebra~\cite{orozco2024quantum,van2026fast,stefanescu2025robust}. 
Such Hamiltonians enable the design of time-dependent parent operators $\Hevo$, governed by the von Neumann equation $\dot{\cal K}(t)=-i[H(t),\Hevo]$, with instantaneous eigenstates following the time-dependent Schr\"odinger equation, even though the explicit construction of such eigenstates is numerically intractable due to the exponentially large Hilbert space~\cite{lewis1967classical,gungordu2012dynamical}.

While there are numerous technical details that distinguish different physical platforms of quantum technological devices, the {\em two-body} interaction Hamiltonian
\be
H_V=g\sum_{j=1}^{n-1}X_jX_{j+1}+Y_jY_{j+1}\ ,
\label{eq:XYinteraction}
\ee
of an open chain of $n$ spins, with an interaction constant $g$ and a nearest-neighbor interaction in terms of Pauli-operators $X$ and $Y$ is a unifying feature of devices comprized of superconducting qubits~\cite{guo2021observation,ye2019propagation}, Rydberg atoms~\cite{barredo2015coherent,scholl2022microwave}, ultracold atoms in optical lattices~\cite{jepsen2020spin,jepsen2021transverse}, and ultracold molecules in optical tweezer arrays~\cite{truppe2017molecules,ruttley2025long}. 

This interaction conserves the total magnetization $M_Z=\sum_{j=1}^nZ_j$, {\it i.e.} $[M_Z,H_V]=0$,
but this conservation can be broken with additional single-qubit terms added to the system Hamiltonian.
Since the terms $Z_j$ also preserve the magnetization, it is necessary to include additional $X_j$ or $Y_j$ terms in the system Hamiltonian in order to reach parent Hamiltonians beyond those that conserve $M_Z$.

The Lie algebra generated by the interaction terms $X_jX_{j+1}+Y_jY_{j+1}$ and additional single qubit $Z_j$ is a subset of the Lie algebra of the transverse Ising model~\cite{wiersema2024classification,aguilar2024classification}, and it scales quadratically in terms of the number of spins (Eq.~\eqref{eq:liealgebrasusbet} of the End Matter).
Adding additional $X_j$ or $Y_j$ terms typically results in an exponentially large Lie algebra, with the notable exception of the operators $X_1$, $Y_1$, $X_n$ and $Y_n$~\cite{orozco2024quantum,van2026fast}.
The Lie algebra generated by the interaction terms $X_jX_{j+1}+Y_jY_{j+1}$, single qubit $Z_j$ terms and general single-qubit terms for qubits $1$ and $n$ is still only quadratically large, and it permits to define system Hamiltonians that do not preserve the total magnetization or any other quantity (see End Matter).

The following analysis is thus based on the system Hamiltonian
\be
H(t)=
h_0(t)X_1+
\sum_{j=1}^{n}h_j(t)Z_j+
h_{n+1}(t)X_n+H_V \ ,
\label{eq:systemHam}
\ee
and a target parent many-body Hamiltonian $\Htarget$ to which a target ground state, with desired topological properties, is an eigenstate to $\Htarget$.
With an initial separable state as an eigenstate to an initial Hamiltonian $\Hinit$, we seek time-dependent functions $h_j(t)$ ($j\in[0,n+1]$) such that evolution under $H(t)$ maps $\Hinit$ to $\Hevofinal$, and $\Hevofinal$ coincides with $\Htarget$ at time $T$, up to a tolerable deviation. 
The same dynamics therefore map the initial separable state to the target ground state.

\paragraph*{Target parent Hamiltonian.}
\begin{table}
\begin{center}
\begin{tabular}{||c || c| c| c |c| c| c |c|}
 \hline
$\stab{S}{j}{\alpha}$
& 1 & 2 & 3 & 4 & 5 & 6 &7\\  
\hline\hline
$\stab{X}{1}{3}$ & $\circleX$ & $\circleZ$ & $\circleZ$ & $\circleX$ & $\circleI$ & $\circleI$ & $\circleI$ \\ 
 \hline
 $\stab{X}{2}{3}$ & $\circleI$ & $\circleX$ & $\circleZ$ & $\circleZ$ & $\circleX$ & $\circleI$ & $\circleI$ \\ 
 \hline
 $\stab{X}{3}{3}$  & $\circleI$& $\circleI$ & $\circleX$ & $\circleZ$ & $\circleZ$ & $\circleX$ & $\circleI$ \\ 
 \hline
 $\stab{X}{4}{3}$   & $\circleI$ & $\circleI$& $\circleI$ & $\circleX$ & $\circleZ$ & $\circleZ$ & $\circleX$\\ 
 \hline
 $\stab{X}{5}{3}$  & $\circleX$ & $\circleI$ & $\circleI$ & $\circleI$ & $\circleX$ & $\circleZ$ & $\circleZ$\\ 
 \hline
$\stab{X}{6}{3}$  & $\circleZ$ & $\circleX$ & $\circleI$ & $\circleI$ & $\circleI$ & $\circleX$ & $\circleZ$\\ 
 \hline
$\stab{X}{7}{3}$  & $\circleZ$ & $\circleZ$ & $\circleX$ & $\circleI$& $\circleI$ & $\circleI$ & $\circleX$\\ 
 \hline
\end{tabular}
\hspace{.2cm}
\begin{tabular}{||c || c| c| c |c| c| c |c|} 
 \hline
$\stab{S}{j}{\alpha}$ & 1 & 2 & 3 & 4 & 5 & 6 &7\\  
\hline\hline
$\stab{X}{1}{3}$ & $\circleX$ & $\circleZ$ & $\circleZ$ & $\circleX$ & $\circleI$ & $\circleI$ & $\circleI$ \\ 
 \hline
 $\stab{X}{2}{3}$ & $\circleI$ & $\circleX$ & $\circleZ$ & $\circleZ$ & $\circleX$ & $\circleI$ & $\circleI$ \\ 
 \hline
 $\stab{X}{3}{3}$  & $\circleI$& $\circleI$ & $\circleX$ & $\circleZ$ & $\circleZ$ & $\circleX$ & $\circleI$ \\ 
 \hline
 $\stab{X}{4}{3}$   & $\circleI$ & $\circleI$& $\circleI$ & $\circleX$ & $\circleZ$ & $\circleZ$ & $\circleX$\\ 
 \hline
 $\stab{Y}{1}{4}$  & $\circleY$ & $\circleZ$ & $\circleZ$ & $\circleZ$ & $\circleY$ & $\circleI$ & $\circleI$\\ 
 \hline
$\stab{Y}{2}{4}$  & $\circleI$ & $\circleY$ & $\circleZ$ & $\circleZ$ & $\circleZ$ & $\circleY$ & $\circleI$\\ 
 \hline
$\stab{Y}{3}{4}$  & $\circleI$ & $\circleI$ & $\circleY$ & $\circleZ$& $\circleZ$ & $\circleZ$ & $\circleY$\\ 
 \hline
\end{tabular}
\caption{
Generalized cluster interactions $\stab{S}{j}{\alpha}$ (Eq.~\eqref{eq:stabilizer}) for a chain of $7$ spins.
The left table depicts the $7$ cluster interactions $\stab{X}{j}{3}$ of this chain.
The operators $\stab{X}{j}{3}$ with $j\in[1,4]$ are elements of the Lie algebra of the system (Eq.~\eqref{eq:systemHam}), but the operators $\stab{X}{j}{3}$ with $j\in[5,7]$ are not; they correspond to PBC.\\
The operators $\stab{X}{j}{3}$ with $j\in[5,7]$ are replaced by the operators $\stab{Y}{j}{4}$ with $j\in[1,3]$ in the right table.
The operators $\stab{Y}{j}{4}$ are elements of the Lie algebra of the system with OBC.}
\label{tab:cluster}
\end{center}
\end{table}

Characterized by the stabilizer interactions
\begin{equation}
\stab{S}{j}{\alpha}=
S_j\Bigl(\prod _{k=1}^{\alpha-1}Z_{j+k}\Bigr)S_{j+\alpha} \ ,
\label{eq:stabilizer}
\end{equation}
with site indices taken modulo $n$ and $\alpha$ denoting the cluster interaction range, the primary target of this work is the generalized cluster parent Hamiltonian
\begin{equation}
    \stb{\mathcal{K}}{\alpha}{m} = -\sum_{j=1}^m \stab{X}{j}{\alpha} \ ,
\label{eq:generalizedcluster}
\end{equation}
with $m=n-\alpha$ for OBC, and $m=n$ for PBC.
The stabilizers with $m>n-\alpha$ are interactions that wrap across the boundaries of the chain (as exemplified in Tab.~\ref{tab:cluster}),
and it is exactly those stabilizers that are missing in OBC.
Given these missing stabilizers, the ground state of $\HOBC$ is $2^\alpha$-fold degenerate.
Under PBC, the ground state is non-degenerate for even $\alpha$.
For odd $\alpha$, it is two-fold degenerate stemming from the fact that the product of all stabilizers $\stab{X}{j}{\alpha}$ yields the identity, making one stabilizer redundant.

While including boundary $X$ drives in the system Hamiltonian is sufficient to realize the cluster interactions with OBC, this fails to produce the PBC interactions;
{\it i.e.} the terms $\stab{X}{j}{\alpha}$ ($j\in [1,n-\alpha]$) are elements of the Lie algebra generated by the system, whereas $\stab{X}{j}{\alpha}$ ($j\in [n-\alpha +1,n]$) are not.
Extending the driving scheme with the goal to realize the missing stabilizers will nearly unavoidably result in an exponentially large Lie algebra, and thus push numerically exact control schemes far out of reach.
The situation can, however, be salvaged by noting that the ground state of the parent Hamiltonian $\HPBC$ (Eq.~\eqref{eq:generalizedcluster}) is an eigenstate to the total parity ${\bf Z}=\prod_{k=1}^nZ_k$ for even $\alpha$ and that a unique ground state within the degenerate subspace for odd $\alpha$ can be defined in terms of the parity ${\bf Z}$, {\it i.e.} with the Hamiltonian $\HPBC + {\bf Z}$.
Since the total parity ${\bf Z}$ commutes with all the stabilizers $\stab{S}{j}{\alpha}$ (Eq.~\eqref{eq:stabilizer}), any such stabilizer in a parent Hamiltonian can by replaced by ${\bf Z}\stab{S}{j}{\alpha}$ 
if an eigenstate to ${\bf Z}$ is targeted.

Crucially, multiplying the parity operator ${\bf Z}$ with a stabilizer
with interactions that wrap across the boundaries of the chain, {\it i.e.} $\stab{X}{n-\alpha+j}{\alpha}$ with $j\in [1,\alpha]$, and thus is not part of the Lie algebra, yields the operator
\begin{equation}
    {\bf Z} \stab{X}{n-\alpha+j}{\alpha}= -\stab{Y}{j}{n-\alpha}\ .
\end{equation}
This operator, however, only has interactions within the chain, and is part of the Lie algebra. 
This is exemplified for the case $n=7$ and $\alpha=3$
in Tab.~\ref{tab:cluster}.

The parent Hamiltonian
\begin{equation}
\HPBCxy = -\sum_{j=1} ^ {n-\alpha} \stab{X}{j}{\alpha} + s\sum_{j=1}^\alpha \stab{Y}{j}{n-\alpha} \ ,
    \label{eq:XYtarget}
\end{equation}
thus has the same ground state as $\HPBC$ (Eq.~\eqref{eq:generalizedcluster}) for even $\alpha$ and $\HPBC+\mathbf{Z}$ for odd $\alpha$, for an appropriate choice of parameter $s=\pm1$, and is comprised only of elements of the quadratically scaling Lie algebra. 
In the End Matter, we show that this requires $s=+1$ when both $n$ and $\alpha$ are even, and $s=-1$ otherwise.

The goal of designing numerically exact control schemes is thus achieved by evolving the initial Hamiltonian $\Hinit$, which is additionally comprised of elements of the quadratically scaling Lie algebra, towards the target parent Hamiltonians $\HOBC$ for OBC and $\HPBCxy$ for PBC where, for this specific Lie algebra, the initial Hamiltonian has the form
\begin{equation}
    \Hinit = c_0\mathbf{Z} + \sum_{j=1}^nc_jZ_j \ ,
    \label{eq:Kinit}
\end{equation}
with parameters $c_j$ determined along with the functions $h_j(t)$, Eq.~\eqref{eq:systemHam} (cf.~\cite{van2026fast}).
Crucial to the practical implementation of the framework, the eigenstates of $\Hinit$ (which can be found efficiently) are product states in the computational basis, and thus can be prepared with little effort.

\paragraph*{Results.} \label{sec:results}

\begin{figure}[t]
    \centering
    \includegraphics[width=\linewidth]{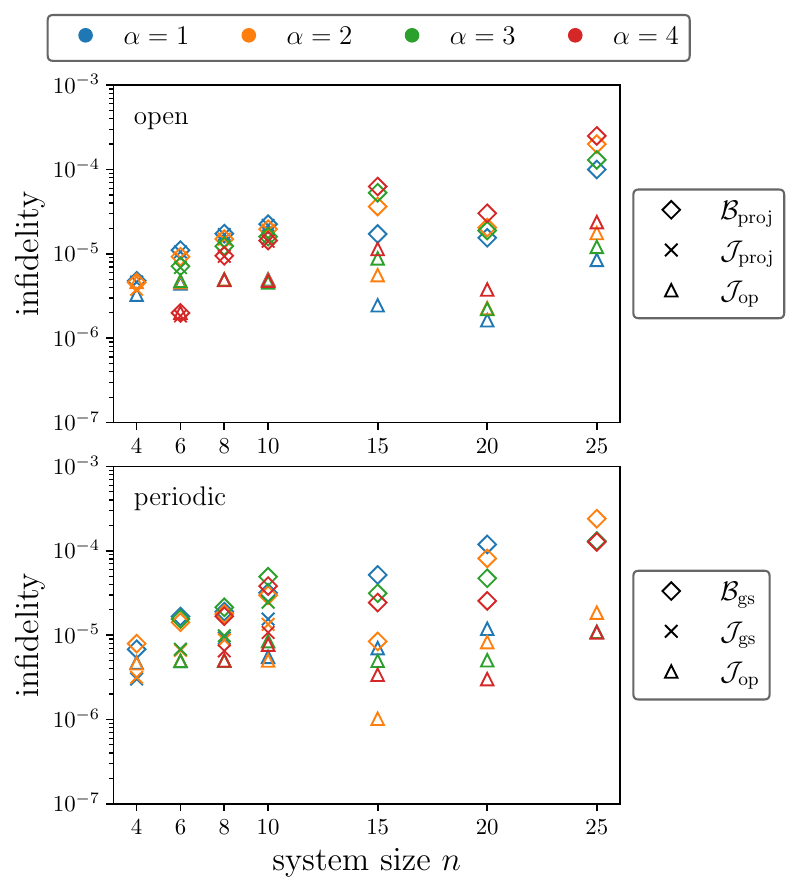}
    \caption{Infidelities as functions of system size $n$ for ground state preparation of the OBC target $\mathcal{K}_\mathbf{T}=\HOBC$ (top panel; Eq.~\eqref{eq:generalizedcluster}) and the PBC target $\mathcal{K}_\mathbf{T}=\HPBCxy$ (bottom panel; Eq.~\eqref{eq:XYtarget}), for cluster interaction ranges $\alpha\in[1,4]$. 
    Triangles show the operator infidelity $\mathcal{J}_\mathrm{op}=\mathcal{J}(\Hevofinal,\mathcal{K}_\mathbf{T})$ (Eq.~\eqref{eq:opJ}). 
    Crosses show the infidelity between the projectors onto the ground spaces of $\Hevofinal$ and $\mathcal{K}_\mathbf{T}$, $\mathcal{J}_\mathrm{proj}$ (Eq.~\eqref{eq:projJ}), in the degenerate OBC case, and the infidelity between the ground states of $\Hevofinal$ and $\mathcal{K}_\mathbf{T}$, $\mathcal{J}_\mathrm{gs}$ (Eq.~\eqref{eq:gsJ}), in the non-degenerate PBC case, while diamonds show their respective upper bounds, $\mathcal{B}_\mathrm{proj}$ (Eq.~\eqref{eq:projBound}) and $\mathcal{B}_\mathrm{gs}$ (Eq.~\eqref{eq:gsbound}). 
    With a typical operator infidelity of $\mathcal{O}(10^{-5})$, the bound shows that the corresponding ground state infidelity does not exceed the remarkably low value of $\mathcal{O}(10^{-4})$.
    }
    \label{fig:infidelity}
\end{figure}

\begin{figure}[t]
    \centering
    \includegraphics[width=\linewidth]{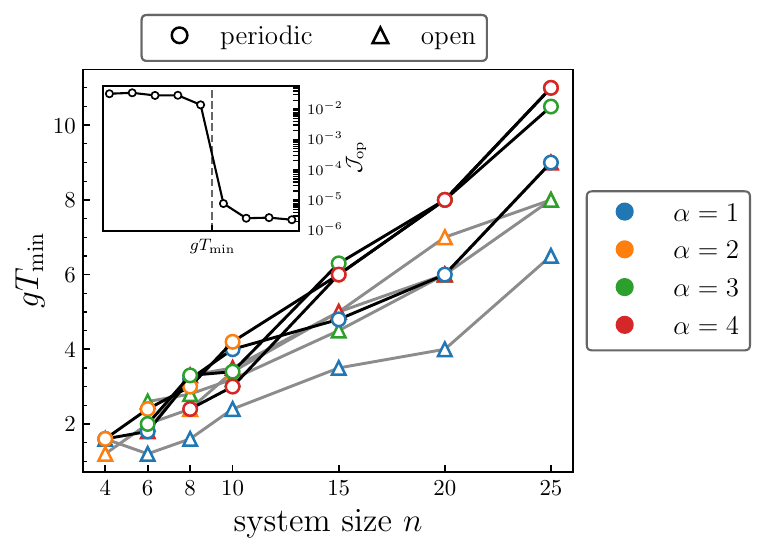}
    \caption{The minimum required duration $gT_\mathrm{min}$, in terms of the interaction strength $g$ of the system Hamiltonian Eq.~\eqref{eq:systemHam}, as a function of system size $n$.
    The triangle points represent target $\mathcal{K}_\mathbf{T}=\HOBC$ (Eq.~\eqref{eq:generalizedcluster}) (OBC), and the circle points represent target $\mathcal{K}_\mathbf{T}=\HPBCxy$ (Eq.~\eqref{eq:XYtarget}) (PBC), for cluster interaction ranges $\alpha \in[1,4]$.
    The minimum duration appears to scale approximately linear with the system size and, for a given system size, the minimum duration does not vary considerably with the cluster interaction range $\alpha$. 
    While the minimum duration does increase for the PBC target Hamiltonian, the duration is still comparable to that of the OBC target Hamiltonian.
    The top left inset shows an example of operator infidelity $\mathcal{J}_\mathrm{op}$ as a function of time, to demonstrate how the minimum duration $gT_\mathrm{min}$ is determined.
    }
    \label{fig:minimumtime}
\end{figure}

The following analysis is thus based on the target parent Hamiltonians $\Htarget=\HOBC$ (Eq.~\eqref{eq:generalizedcluster}) (OBC) and $\Htarget=\HPBCxy$ (Eq.~\eqref{eq:XYtarget}) (PBC).
The optimal time-dependencies $h_j(t)$ ($j\in[0,n+1]$) of the system Hamiltonian (Eq.~\eqref{eq:systemHam}) are determined by the minimization of the deviation of the evolved operator $\Hevofinal$, from the target $\Htarget$,
quantified by the operator infidelity $\mathcal{J}_\mathrm{op}=\mathcal{J}(\Hevofinal,\mathcal{K}_\mathbf{T})$.
The specific function $\mathcal{J}$ is given in Eq.~\eqref{eq:opJ} of the End Matter, which can be evaluated with polynomial effort owing to the small Lie algebra.

While a minimal value of $\mathcal{J}_\text{op}$ indeed achieves the goal of ground state preparation~\cite{orozco2024quantum,van2026fast}, the true figure of merit is the infidelity $\mathcal{J}_\mathrm{proj}$ (Eq.~\eqref{eq:projJ}) between the projectors onto the ground spaces of $\Hevofinal$ and $\HOBC$ for the degenerate case (OBC), and the pure state infidelity $\mathcal{J}_\mathrm{gs}$ (Eq.~\eqref{eq:gsJ})
between the ground states of $\Hevofinal$ and $\HPBCxy$ for the non-degenerate case (PBC).
The evaluation of both quantities, however, generally has an effort that scales with the size of the Hilbert space and thus exponentially.
The upper bounds to these quantities, $\mathcal{B}_\mathrm{proj}$ (Eq.~\eqref{eq:projBound}) and $\mathcal{B}_\mathrm{gs}$ (Eq.~\eqref{eq:gsbound}), however, can be evaluated with polynomially scaling effort (cf.~\cite{orozco2024quantum}).

Fig.~\ref{fig:infidelity} shows these quantities as a function of system size $n$, for the $2^\alpha$-fold degenerate target (OBC) in the top panel, and for the non-degenerate target (PBC) in the bottom panel, for cluster interaction ranges $\alpha\in [1,4]$.
The operator infidelity $\mathcal{J}_\mathrm{op}$ is depicted by triangles, the corresponding $\mathcal{J}_\mathrm{proj}$ and $\mathcal{J}_\mathrm{gs}$ infidelities are depicted by crosses, and their upper bounds are depicted by diamonds.

The results are comparable for the OBC and PBC cases. 
In all instances, operator infidelities of $\mathcal{O}(10^{-5})$ correspond to remarkably low upper bounds on the ground state infidelity of at most $\mathcal{O}(10^{-4})$. 
For a fixed system size, these bounds also depend only weakly on the interaction range $\alpha$; for example, all diamond markers at $n=25$ lie within a narrow range of infidelities.

To realize states with these infidelities, one requires a minimum control duration $gT_\mathrm{min}$, in terms of the interaction strength $g$ of the system Hamiltonian (Eq.~\eqref{eq:XYinteraction}).
While the minimum duration generally does not have analytical expressions, it can be determined fairly accurately via the minimization of the infidelity $\mathcal{J}_\mathrm{op}$ over a range of control durations~\cite{van2026fast}.
As the duration is increased, there is gradual improvement to the infidelity until a specific duration where the infidelity drops several orders of magnitude over a short interval of time.
This specific duration is recognized as the minimum duration, 
with an example of this shown in the top left inset of Fig.~\ref{fig:minimumtime}.

Fig.~\ref{fig:minimumtime} also shows $gT_\text{min}$ as a function of system size $n$ for both the OBC target with points depicted by triangles, and the PBC target with points depicted by circles, for cluster interaction ranges $\alpha\in [1,4]$. 
Despite the limited reliability of numerical optimization, a linear growth of $gT_\mathrm{min}$ with system size $n$ is rather evident for both the open and periodic case, as well as the observation that $gT_\mathrm{min}$ does not vary considerably with the cluster interaction range $\alpha$.
This scaling is remarkably weak, only requiring a small percentage of the full coherence budget of modern quantum simulators~\cite{lu2026probing} at large system sizes, such as $n=25$.
While the minimum duration $gT_\mathrm{min}$ in order to realize the parent Hamiltonian for the periodic case increases, due to the requirement to numerically realize the additional effective interactions $\stab{Y}{j}{n-\alpha}$, strikingly, the durations are still largely comparable to the minimum duration to realize the parent Hamiltonian for the open case.

\paragraph*{Discussion.} 
With quantum states defined via mutually commuting operators in various applications, the present techniques for state preparation with properties that seem incompatible with native system characteristics is applicable to the full spectrum of problems defined in terms of the stabilizer formalism.

While Hamiltonians associated with a small Lie algebra can fundamentally only induce the dynamics associated with the corresponding Lie group, the set of parent Hamiltonians for state preparation is not strictly limited by the Lie algebra,
since a stabilizer lying outside the small Lie algebra may be replaced by a suitable product of stabilizers that belongs to the Lie algebra while preserving the relevant stabilizer eigenstate.

Given a non-degenerate initial parent Hamiltonian $\Hinit$ the present state preparation protocol also applies directly to excited states once an optimized driving pattern is identified.
As such, the 
state preparation techniques described here can enable access to a broad range of correlated quantum states that might be prohibitively difficult to reach with previously existing methods.

The moderate scaling of the required evolution time with both system size and interaction complexity hints at the potential of analog protocols for scalable quantum simulation and computation, as well as for the preparation of resource states for subsequent measurement-based quantum computation. 
More broadly, our results demonstrate that the geometry of a quantum device need not fundamentally restrict the class of states that can be prepared.

\paragraph*{Acknowledgements.}
This work was supported by the U.K. Engineering and Physical Sciences Research Council (EPSRC DTP - EP/W524323/1) and by the National Natural Science
Foundation of China (Grant No. 12474214) and by Quantum Science and Technology-National Science and Technology Major Project
(No. 2024ZD0301800).
Numerical simulations and optimization routines were performed on the Imperial HPC cluster.
\paragraph*{Data availability.}
The optimal control solutions used in this paper are
available without restriction~\cite{van_lomwel_data}.

\bibliography{library_PRL_XXYY.bib}

@article{georgescu2014quantum,
  title = {Quantum simulation},
  author = {Georgescu, I. M. and Ashhab, S. and Nori, Franco},
  journal = {Rev. Mod. Phys.},
  volume = {86},
  issue = {1},
  pages = {153--185},
  numpages = {33},
  year = {2014},
  month = {Mar},
  publisher = {American Physical Society},
  doi = {10.1103/RevModPhys.86.153},
  url = {https://link.aps.org/doi/10.1103/RevModPhys.86.153}
}

@article{altman2021quantum,
  title = {Quantum Simulators: Architectures and Opportunities},
  author = {Altman, Ehud and Brown, Kenneth R. and Carleo, Giuseppe and Carr, Lincoln D. and Demler, Eugene and Chin, Cheng and DeMarco, Brian and Economou, Sophia E. and Eriksson, Mark A. and Fu, Kai-Mei C. and others},
  journal = {PRX Quantum},
  volume = {2},
  issue = {1},
  pages = {017003},
  numpages = {19},
  year = {2021},
  month = {Feb},
  publisher = {American Physical Society},
  doi = {10.1103/PRXQuantum.2.017003},
  url = {https://link.aps.org/doi/10.1103/PRXQuantum.2.017003}
}

@article{daley2022practical,
  title={Practical quantum advantage in quantum simulation},
  author={Daley, Andrew J and Bloch, Immanuel and Kokail, Christian and Flannigan, Stuart and Pearson, Natalie and Troyer, Matthias and Zoller, Peter},
  journal={Nature},
  volume={607},
  number={7920},
  pages={667--676},
  year={2022},
  publisher={Nature Publishing Group UK London},
  url={https://doi.org/10.1038/s41586-022-04940-6}
}

@article{scholl2022microwave,
  title = {Microwave Engineering of Programmable $XXZ$ Hamiltonians in Arrays of Rydberg Atoms},
  author = {Scholl, P. and Williams, H. J. and Bornet, G. and Wallner, F. and Barredo, D. and Henriet, L. and Signoles, A. and Hainaut, C. and Franz, T. and Geier, S. and Tebben, A. and Salzinger, A. and Z\"urn, G. and Lahaye, T. and Weidem\"uller, M. and Browaeys, A.},
  journal = {PRX Quantum},
  volume = {3},
  issue = {2},
  pages = {020303},
  numpages = {10},
  year = {2022},
  month = {Apr},
  publisher = {American Physical Society},
  doi = {10.1103/PRXQuantum.3.020303},
  url = {https://link.aps.org/doi/10.1103/PRXQuantum.3.020303}
}

@article{periwal2021programmable,
  title = {Programmable interactions and emergent geometry in an array of atom clouds},
  volume = {600},
  ISSN = {1476-4687},
  url = {http://dx.doi.org/10.1038/s41586-021-04156-0},
  DOI = {10.1038/s41586-021-04156-0},
  number = {7890},
  journal = {Nature},
  publisher = {Springer Science and Business Media LLC},
  author = {Periwal,  Avikar and Cooper,  Eric S. and Kunkel,  Philipp and Wienand,  Julian F. and Davis,  Emily J. and Schleier-Smith,  Monika},
  year = {2021},
  month = Dec,
  pages = {630–635}
}

@article{campostrini2014finite,
  title = {Finite-size scaling at quantum transitions},
  author = {Campostrini, Massimo and Pelissetto, Andrea and Vicari, Ettore},
  journal = {Phys. Rev. B},
  volume = {89},
  issue = {9},
  pages = {094516},
  numpages = {22},
  year = {2014},
  month = {Mar},
  publisher = {American Physical Society},
  doi = {10.1103/PhysRevB.89.094516},
  url = {https://link.aps.org/doi/10.1103/PhysRevB.89.094516}
}

@article{lahtinen2015realizing,
  title = {Realizing All $so(N{)}_{1}$ Quantum Criticalities in Symmetry Protected Cluster Models},
  author = {Lahtinen, Ville and Ardonne, Eddy},
  journal = {Phys. Rev. Lett.},
  volume = {115},
  issue = {23},
  pages = {237203},
  numpages = {5},
  year = {2015},
  month = {Dec},
  publisher = {American Physical Society},
  doi = {10.1103/PhysRevLett.115.237203},
  url = {https://link.aps.org/doi/10.1103/PhysRevLett.115.237203}
}

@article{ohta2016topological,
  title = {Topological and dynamical properties of a generalized cluster model in one dimension},
  author = {Ohta, Takumi and Tanaka, Shu and Danshita, Ippei and Totsuka, Keisuke},
  journal = {Phys. Rev. B},
  volume = {93},
  issue = {16},
  pages = {165423},
  numpages = {15},
  year = {2016},
  month = {Apr},
  publisher = {American Physical Society},
  doi = {10.1103/PhysRevB.93.165423},
  url = {https://link.aps.org/doi/10.1103/PhysRevB.93.165423}
}

@article{verresen2017one,
  title = {One-dimensional symmetry protected topological phases and their transitions},
  author = {Verresen, Ruben and Moessner, Roderich and Pollmann, Frank},
  journal = {Phys. Rev. B},
  volume = {96},
  issue = {16},
  pages = {165124},
  numpages = {23},
  year = {2017},
  month = {Oct},
  publisher = {American Physical Society},
  doi = {10.1103/PhysRevB.96.165124},
  url = {https://link.aps.org/doi/10.1103/PhysRevB.96.165124}
}

@article{raussendorf2001one,
  title = {A One-Way Quantum Computer},
  author = {Raussendorf, Robert and Briegel, Hans J.},
  journal = {Phys. Rev. Lett.},
  volume = {86},
  issue = {22},
  pages = {5188--5191},
  numpages = {0},
  year = {2001},
  month = {May},
  publisher = {American Physical Society},
  doi = {10.1103/PhysRevLett.86.5188},
  url = {https://link.aps.org/doi/10.1103/PhysRevLett.86.5188}
}

@article{else2012symmetry,
  title = {Symmetry-Protected Phases for Measurement-Based Quantum Computation},
  author = {Else, Dominic V. and Schwarz, Ilai and Bartlett, Stephen D. and Doherty, Andrew C.},
  journal = {Phys. Rev. Lett.},
  volume = {108},
  issue = {24},
  pages = {240505},
  numpages = {5},
  year = {2012},
  month = {Jun},
  publisher = {American Physical Society},
  doi = {10.1103/PhysRevLett.108.240505},
  url = {https://link.aps.org/doi/10.1103/PhysRevLett.108.240505}
}

@article{son2011topological,
   title = {Topological order in 1D Cluster state protected by symmetry},
  volume = {11},
  ISSN = {1573-1332},
  url = {http://dx.doi.org/10.1007/s11128-011-0346-7},
  DOI = {10.1007/s11128-011-0346-7},
  number = {6},
  journal = {Quantum Information Processing},
  publisher = {Springer Science and Business Media LLC},
  author = {Son,  W. and Amico,  L. and Vedral,  V.},
  year = {2011},
  month = Dec,
  pages = {1961–1968}
   }

@article{smacchia2011statistical,
  title = {Statistical mechanics of the cluster Ising model},
  author = {Smacchia, Pietro and Amico, Luigi and Facchi, Paolo and Fazio, Rosario and Florio, Giuseppe and Pascazio, Saverio and Vedral, Vlatko},
  journal = {Phys. Rev. A},
  volume = {84},
  issue = {2},
  pages = {022304},
  numpages = {12},
  year = {2011},
  month = {Aug},
  publisher = {American Physical Society},
  doi = {10.1103/PhysRevA.84.022304},
  url = {https://link.aps.org/doi/10.1103/PhysRevA.84.022304}
}

@article{perez2008string,
  title = {String Order and Symmetries in Quantum Spin Lattices},
  author = {P\'erez-Garc\'{\i}a, D. and Wolf, M. M. and Sanz, M. and Verstraete, F. and Cirac, J. I.},
  journal = {Phys. Rev. Lett.},
  volume = {100},
  issue = {16},
  pages = {167202},
  numpages = {4},
  year = {2008},
  month = {Apr},
  publisher = {American Physical Society},
  doi = {10.1103/PhysRevLett.100.167202},
  url = {https://link.aps.org/doi/10.1103/PhysRevLett.100.167202}
}

@article{gong2019genuine,
  title = {Genuine 12-Qubit Entanglement on a Superconducting Quantum Processor},
  author = {Gong, Ming and Chen, Ming-Cheng and Zheng, Yarui and Wang, Shiyu and Zha, Chen and Deng, Hui and Yan, Zhiguang and Rong, Hao and Wu, Yulin and Li, Shaowei and others},
  journal = {Phys. Rev. Lett.},
  volume = {122},
  issue = {11},
  pages = {110501},
  numpages = {5},
  year = {2019},
  month = {Mar},
  publisher = {American Physical Society},
  doi = {10.1103/PhysRevLett.122.110501},
  url = {https://link.aps.org/doi/10.1103/PhysRevLett.122.110501}
}

@article{bluvstein2022quantum,
  title = {A quantum processor based on coherent transport of entangled atom arrays},
  volume = {604},
  ISSN = {1476-4687},
  url = {http://dx.doi.org/10.1038/s41586-022-04592-6},
  DOI = {10.1038/s41586-022-04592-6},
  number = {7906},
  journal = {Nature},
  publisher = {Springer Science and Business Media LLC},
  author = {Bluvstein,  Dolev and Levine,  Harry and Semeghini,  Giulia and Wang,  Tout T. and Ebadi,  Sepehr and Kalinowski,  Marcin and Keesling,  Alexander and Maskara,  Nishad and Pichler,  Hannes and Greiner,  Markus and Vuletić,  Vladan and Lukin,  Mikhail D.},
  year = {2022},
  month = Apr,
  pages = {451–456}
}

@article{tan2023realizing,
  title = {Realizing symmetry-protected topological phases in a spin-1/2 chain with next-nearest-neighbor hopping on superconducting qubits},
  author = {Tan, Adrian T. K. and Sun, Shi-Ning and Tazhigulov, Ruslan N. and Chan, Garnet Kin-Lic and Minnich, Austin J.},
  journal = {Phys. Rev. A},
  volume = {107},
  issue = {3},
  pages = {032614},
  numpages = {9},
  year = {2023},
  month = {Mar},
  publisher = {American Physical Society},
  doi = {10.1103/PhysRevA.107.032614},
  url = {https://link.aps.org/doi/10.1103/PhysRevA.107.032614}
}

@article{iqbal2024topological,
  title={Topological order from measurements and feed-forward on a trapped ion quantum computer},
  author={Iqbal, Mohsin and Tantivasadakarn, Nathanan and Gatterman, Thomas M and Gerber, Justin A and Gilmore, Kevin and Gresh, Dan and Hankin, Aaron and Hewitt, Nathan and Horst, Chandler V and Matheny, Mitchell and others},
  journal={Communications Physics},
  volume={7},
  number={1},
  pages={205},
  year={2024},
  publisher={Nature Publishing Group UK London},
  url={http://dx.doi.org/10.1038/s42005-024-01698-3}
}

@article{kuznetsova2012yelin,
  title = {Cluster-state generation using van der Waals and dipole-dipole interactions in optical lattices},
  author = {Kuznetsova, Elena and Bragdon, T. and C\^ot\'e, Robin and Yelin, S. F.},
  journal = {Phys. Rev. A},
  volume = {85},
  issue = {1},
  pages = {012328},
  numpages = {20},
  year = {2012},
  month = {Jan},
  publisher = {American Physical Society},
  doi = {10.1103/PhysRevA.85.012328},
  url = {https://link.aps.org/doi/10.1103/PhysRevA.85.012328}
}

@article{son2011quantum,
  title = {Quantum phase transition between cluster and antiferromagnetic states},
  volume = {95},
  ISSN = {1286-4854},
  url = {http://dx.doi.org/10.1209/0295-5075/95/50001},
  DOI = {10.1209/0295-5075/95/50001},
  number = {5},
  journal = {EPL (Europhysics Letters)},
  publisher = {IOP Publishing},
  author = {Son,  W. and Amico,  L. and Fazio,  R. and Hamma,  A. and Pascazio,  S. and Vedral,  V.},
  year = {2011},
  month = Aug,
  pages = {50001}
}

@article{seifnashri2024cluster,
  title = {Cluster State as a Noninvertible Symmetry-Protected Topological Phase},
  author = {Seifnashri, Sahand and Shao, Shu-Heng},
  journal = {Phys. Rev. Lett.},
  volume = {133},
  issue = {11},
  pages = {116601},
  numpages = {7},
  year = {2024},
  month = {Sep},
  publisher = {American Physical Society},
  doi = {10.1103/PhysRevLett.133.116601},
  url = {https://link.aps.org/doi/10.1103/PhysRevLett.133.116601}
}

@article{zhang2022digital,
  title = {Digital quantum simulation of Floquet symmetry-protected topological phases},
  volume = {607},
  ISSN = {1476-4687},
  url = {http://dx.doi.org/10.1038/s41586-022-04854-3},
  DOI = {10.1038/s41586-022-04854-3},
  number = {7919},
  journal = {Nature},
  publisher = {Springer Science and Business Media LLC},
  author = {Zhang,  Xu and Jiang,  Wenjie and Deng,  Jinfeng and Wang,  Ke and Chen,  Jiachen and Zhang,  Pengfei and Ren,  Wenhui and Dong,  Hang and Xu,  Shibo and Gao,  Yu and others},
  year = {2022},
  month = July,
  pages = {468–473}
}

@article{orozco2024quantum,
  title = {Quantum Control without Quantum States},
  author = {Orozco-Ruiz, Modesto and Le, Nguyen H. and Mintert, Florian},
  journal = {PRX Quantum},
  volume = {5},
  issue = {4},
  pages = {040346},
  numpages = {22},
  year = {2024},
  month = {Dec},
  publisher = {American Physical Society},
  doi = {10.1103/PRXQuantum.5.040346},
  url = {https://link.aps.org/doi/10.1103/PRXQuantum.5.040346}
}

@misc{van2026fast,
      title={Fast thermal state preparation beyond native interactions}, 
      author={Alexander van Lomwel and Paul M. Schindler and Modesto Orozco-Ruiz and Marin Bukov and Nguyen H. Le and Florian Mintert},
      year={2026},
      eprint={2601.04810},
      archivePrefix={arXiv},
      primaryClass={quant-ph},
      url={https://arxiv.org/abs/2601.04810}, 
}

@misc{lu2026probing,
      title={Probing Coherent Many-Body Spin Dynamics in a Molecular Tweezer Array Quantum Simulator}, 
      author={Yukai Lu and Connor M. Holland and Callum L. Welsh and Xing-Yan Chen and Lawrence W. Cheuk},
      year={2026},
      eprint={2603.19090},
      archivePrefix={arXiv},
      primaryClass={cond-mat.quant-gas},
      url={https://arxiv.org/abs/2603.19090}, 
}

@article{stefanescu2025robust,
  title = {Robust implicit quantum control of interacting spin chains},
  author = {Stefanescu, Luca and Edwards-Pratt, Louis and O'Connor, Jeremy and Tsegaye, Ezra and Le, Nguyen H. and Mintert, Florian},
  journal = {Phys. Rev. A},
  volume = {112},
  issue = {1},
  pages = {012609},
  numpages = {12},
  year = {2025},
  month = {Jul},
  publisher = {American Physical Society},
  doi = {10.1103/zmcn-ddkj},
  url = {https://link.aps.org/doi/10.1103/zmcn-ddkj}
}

@article{wiersema2024classification,
  title={Classification of dynamical Lie algebras of 2-local spin systems on linear, circular and fully connected topologies},
  author={Wiersema, Roeland and K{\"o}kc{\"u}, Efekan and Kemper, Alexander F and Bakalov, Bojko N},
  journal={npj Quantum Information},
  volume={10},
  number={1},
  pages={110},
  year={2024},
  publisher={Nature Publishing Group UK London},
url = {https://doi.org/10.1038/s41534-024-00900-2}
}

@misc{aguilar2024classification,
      title={Full classification of Pauli Lie algebras}, 
      author={Gerard Aguilar and Simon Cichy and Jens Eisert and Lennart Bittel},
      year={2024},
      eprint={2408.00081},
      archivePrefix={arXiv},
      primaryClass={quant-ph},
      url={https://arxiv.org/abs/2408.00081}, 
}

@article{petiziol2024nonperturbative,
  title = {Nonperturbative Floquet engineering of the toric-code Hamiltonian and its ground state},
  author = {Petiziol, Francesco and Wimberger, Sandro and Eckardt, Andr\'e and Mintert, Florian},
  journal = {Phys. Rev. B},
  volume = {109},
  issue = {7},
  pages = {075126},
  numpages = {19},
  year = {2024},
  month = {Feb},
  publisher = {American Physical Society},
  doi = {10.1103/PhysRevB.109.075126},
  url = {https://link.aps.org/doi/10.1103/PhysRevB.109.075126}
}

@article{petiziol2021quantum,
  title = {Quantum Simulation of Three-Body Interactions in Weakly Driven Quantum Systems},
  author = {Petiziol, Francesco and Sameti, Mahdi and Carretta, Stefano and Wimberger, Sandro and Mintert, Florian},
  journal = {Phys. Rev. Lett.},
  volume = {126},
  issue = {25},
  pages = {250504},
  numpages = {6},
  year = {2021},
  month = {Jun},
  publisher = {American Physical Society},
  doi = {10.1103/PhysRevLett.126.250504},
  url = {https://link.aps.org/doi/10.1103/PhysRevLett.126.250504}
}

@article{kitaev2003fault,
   title={Fault-tolerant quantum computation by anyons},
   volume={303},
   ISSN={0003-4916},
   url={http://dx.doi.org/10.1016/S0003-4916(02)00018-0},
   DOI={10.1016/s0003-4916(02)00018-0},
   number={1},
   journal={Annals of Physics},
   publisher={Elsevier BV},
   author={Kitaev, A.Yu.},
   year={2003},
   month=Jan, pages={2–30} }

@article{bombin2006topological,
  title = {Topological Quantum Distillation},
  author = {Bombin, H. and Martin-Delgado, M. A.},
  journal = {Phys. Rev. Lett.},
  volume = {97},
  issue = {18},
  pages = {180501},
  numpages = {4},
  year = {2006},
  month = {Oct},
  publisher = {American Physical Society},
  doi = {10.1103/PhysRevLett.97.180501},
  url = {https://link.aps.org/doi/10.1103/PhysRevLett.97.180501}
}

@article{bravyi2010tradeoffs,
  title = {Tradeoffs for Reliable Quantum Information Storage in 2D Systems},
  author = {Bravyi, Sergey and Poulin, David and Terhal, Barbara},
  journal = {Phys. Rev. Lett.},
  volume = {104},
  issue = {5},
  pages = {050503},
  numpages = {4},
  year = {2010},
  month = {Feb},
  publisher = {American Physical Society},
  doi = {10.1103/PhysRevLett.104.050503},
  url = {https://link.aps.org/doi/10.1103/PhysRevLett.104.050503}
}

@article{wecker2015progress,
  title = {Progress towards practical quantum variational algorithms},
  author = {Wecker, Dave and Hastings, Matthew B. and Troyer, Matthias},
  journal = {Phys. Rev. A},
  volume = {92},
  issue = {4},
  pages = {042303},
  numpages = {10},
  year = {2015},
  month = {Oct},
  publisher = {American Physical Society},
  doi = {10.1103/PhysRevA.92.042303},
  url = {https://link.aps.org/doi/10.1103/PhysRevA.92.042303}
}

@article{malley2016scalable,
  title = {Scalable Quantum Simulation of Molecular Energies},
  author = {O'Malley, P. J. J. and Babbush, R. and Kivlichan, I. D. and Romero, J. and McClean, J. R. and Barends, R. and Kelly, J. and Roushan, P. and Tranter, A. and Ding, N. and others},
  journal = {Phys. Rev. X},
  volume = {6},
  issue = {3},
  pages = {031007},
  numpages = {13},
  year = {2016},
  month = {Jul},
  publisher = {American Physical Society},
  doi = {10.1103/PhysRevX.6.031007},
  url = {https://link.aps.org/doi/10.1103/PhysRevX.6.031007}
}

@article{farrell2024scalable,
  title = {Scalable Circuits for Preparing Ground States on Digital Quantum Computers: The Schwinger Model Vacuum on 100 Qubits},
  author = {Farrell, Roland C. and Illa, Marc and Ciavarella, Anthony N. and Savage, Martin J.},
  journal = {PRX Quantum},
  volume = {5},
  issue = {2},
  pages = {020315},
  numpages = {32},
  year = {2024},
  month = {Apr},
  publisher = {American Physical Society},
  doi = {10.1103/PRXQuantum.5.020315},
  url = {https://link.aps.org/doi/10.1103/PRXQuantum.5.020315}
}

@misc{farhi2014quantum,
      title={A Quantum Approximate Optimization Algorithm}, 
      author={Edward Farhi and Jeffrey Goldstone and Sam Gutmann},
      year={2014},
      eprint={1411.4028},
      archivePrefix={arXiv},
      primaryClass={quant-ph},
      url={https://arxiv.org/abs/1411.4028}, 
}

@article{albash2018adiabatic,
  title = {Adiabatic quantum computation},
  author = {Albash, Tameem and Lidar, Daniel A.},
  journal = {Rev. Mod. Phys.},
  volume = {90},
  issue = {1},
  pages = {015002},
  numpages = {64},
  year = {2018},
  month = {Jan},
  publisher = {American Physical Society},
  doi = {10.1103/RevModPhys.90.015002},
  url = {https://link.aps.org/doi/10.1103/RevModPhys.90.015002}
}

@article{zhou2020quantum,
  title = {Quantum Approximate Optimization Algorithm: Performance, Mechanism, and Implementation on Near-Term Devices},
  author = {Zhou, Leo and Wang, Sheng-Tao and Choi, Soonwon and Pichler, Hannes and Lukin, Mikhail D.},
  journal = {Phys. Rev. X},
  volume = {10},
  issue = {2},
  pages = {021067},
  numpages = {23},
  year = {2020},
  month = {Jun},
  publisher = {American Physical Society},
  doi = {10.1103/PhysRevX.10.021067},
  url = {https://link.aps.org/doi/10.1103/PhysRevX.10.021067}
}

@article{nguyen2023quantum,
  title = {Quantum Optimization with Arbitrary Connectivity Using Rydberg Atom Arrays},
  author = {Nguyen, Minh-Thi and Liu, Jin-Guo and Wurtz, Jonathan and Lukin, Mikhail D. and Wang, Sheng-Tao and Pichler, Hannes},
  journal = {PRX Quantum},
  volume = {4},
  issue = {1},
  pages = {010316},
  numpages = {19},
  year = {2023},
  month = {Feb},
  publisher = {American Physical Society},
  doi = {10.1103/PRXQuantum.4.010316},
  url = {https://link.aps.org/doi/10.1103/PRXQuantum.4.010316}
}

@article{preskill2018quantum,
   title={Quantum Computing in the NISQ era and beyond},
   volume={2},
   ISSN={2521-327X},
   url={http://dx.doi.org/10.22331/q-2018-08-06-79},
   DOI={10.22331/q-2018-08-06-79},
   journal={Quantum},
   publisher={Verein zur Forderung des Open Access Publizierens in den Quantenwissenschaften},
   author={Preskill, John},
   year={2018},
   month=Aug, pages={79} }

@misc{flannigan2022propagation,
      title={Propagation of errors and quantitative quantum simulation with quantum advantage}, 
      author={S. Flannigan and N. Pearson and G. H. Low and A. Buyskikh and I. Bloch and P. Zoller and M. Troyer and A. J. Daley},
      year={2022},
      eprint={2204.13644},
      archivePrefix={arXiv},
      primaryClass={quant-ph},
      url={https://arxiv.org/abs/2204.13644}, 
}

@article{decker2020floquet,
  title = {Floquet Engineering Topological Many-Body Localized Systems},
  author = {Decker, K. S. C. and Karrasch, C. and Eisert, J. and Kennes, D. M.},
  journal = {Phys. Rev. Lett.},
  volume = {124},
  issue = {19},
  pages = {190601},
  numpages = {7},
  year = {2020},
  month = {May},
  publisher = {American Physical Society},
  doi = {10.1103/PhysRevLett.124.190601},
  url = {https://link.aps.org/doi/10.1103/PhysRevLett.124.190601}
}

@article{koyluoglu2025floquet,
  title = {Floquet Engineering of Interactions and Entanglement in Periodically Driven Rydberg Chains},
  author = {Koyluoglu, Nazli Ugur and Maskara, Nishad and Feldmeier, Johannes and Lukin, Mikhail D.},
  journal = {Phys. Rev. Lett.},
  volume = {135},
  issue = {11},
  pages = {113603},
  numpages = {9},
  year = {2025},
  month = {Sep},
  publisher = {American Physical Society},
  doi = {10.1103/5qhh-322q},
  url = {https://link.aps.org/doi/10.1103/5qhh-322q}
}

@article{wu2026engineering,
  title = {Engineering Long-Range and Multibody Interactions via Global Kinetic Constraints},
  author = {Wu, Runmin and Yang, Bing and Claeys, Pieter W. and Zhao, Hongzheng},
  journal = {Phys. Rev. Lett.},
  volume = {136},
  issue = {12},
  pages = {120401},
  numpages = {9},
  year = {2026},
  month = {Mar},
  publisher = {American Physical Society},
  doi = {10.1103/4l1s-kkw7},
  url = {https://link.aps.org/doi/10.1103/4l1s-kkw7}
}

@article{maskara2025programmable,
  title = {Programmable simulations of molecules and materials with reconfigurable quantum processors},
  volume = {21},
  ISSN = {1745-2481},
  url = {http://dx.doi.org/10.1038/s41567-024-02738-z},
  DOI = {10.1038/s41567-024-02738-z},
  number = {2},
  journal = {Nature Physics},
  publisher = {Springer Science and Business Media LLC},
  author = {Maskara,  Nishad and Ostermann,  Stefan and Shee,  James and Kalinowski,  Marcin and McClain Gomez,  Abigail and Araiza Bravo,  Rodrigo and Wang,  Derek S. and Krylov,  Anna I. and Yao,  Norman Y. and Head-Gordon,  Martin and Lukin,  Mikhail D. and Yelin,  Susanne F.},
  year = {2025},
  month = Jan,
  pages = {289–297}
}

@article{greenaway2024variational,
  title={Variational quantum gate optimization at the pulse level},
  author={Greenaway, Sean and Petiziol, Francesco and Hongzheng, Zhao and Mintert, Florian},
  journal={SciPost Physics},
  volume={16},
  number={3},
  pages={082},
  year={2024},
  url={http://dx.doi.org/10.21468/scipostphys.16.3.082}
}

@article{eckardt2015high,
   title={High-frequency approximation for periodically driven quantum systems from a Floquet-space perspective},
   volume={17},
   ISSN={1367-2630},
   url={http://dx.doi.org/10.1088/1367-2630/17/9/093039},
   DOI={10.1088/1367-2630/17/9/093039},
   number={9},
   journal={New Journal of Physics},
   publisher={IOP Publishing},
   author={Eckardt, André and Anisimovas, Egidijus},
   year={2015},
   month=Sept, pages={093039} }

@misc{baumgartner2025hilbert,
      title={Hilbert Space Diffusion in Systems with Approximate Symmetries}, 
      author={Rahel L. Baumgartner and Luca V. Delacrétaz and Pranjal Nayak and Julian Sonner},
      year={2025},
      eprint={2405.19260},
      archivePrefix={arXiv},
      primaryClass={cond-mat.stat-mech},
      url={https://arxiv.org/abs/2405.19260}, 
}

@article{guo2021observation,
  title={Observation of Bloch oscillations and Wannier-Stark localization on a superconducting quantum processor},
  author={Guo, Xue-Yi and Ge, Zi-Yong and Li, Hekang and Wang, Zhan and Zhang, Yu-Ran and Song, Pengtao and Xiang, Zhongcheng and Song, Xiaohui and Jin, Yirong and Lu, Li and others},
  journal={npj Quantum Information},
  volume={7},
  number={1},
  pages={51},
  year={2021},
  publisher={Nature Publishing Group UK London},
  url={http://dx.doi.org/10.1038/s41534-021-00385-3}
}

@article{ye2019propagation,
  title = {Propagation and Localization of Collective Excitations on a 24-Qubit Superconducting Processor},
  author = {Ye, Yangsen and Ge, Zi-Yong and Wu, Yulin and Wang, Shiyu and Gong, Ming and Zhang, Yu-Ran and Zhu, Qingling and Yang, Rui and Li, Shaowei and Liang, Futian and others},
  journal = {Phys. Rev. Lett.},
  volume = {123},
  issue = {5},
  pages = {050502},
  numpages = {6},
  year = {2019},
  month = {Jul},
  publisher = {American Physical Society},
  doi = {10.1103/PhysRevLett.123.050502},
  url = {https://link.aps.org/doi/10.1103/PhysRevLett.123.050502}
}

@article{barredo2015coherent,
  title = {Coherent Excitation Transfer in a Spin Chain of Three Rydberg Atoms},
  author = {Barredo, Daniel and Labuhn, Henning and Ravets, Sylvain and Lahaye, Thierry and Browaeys, Antoine and Adams, Charles S.},
  journal = {Phys. Rev. Lett.},
  volume = {114},
  issue = {11},
  pages = {113002},
  numpages = {5},
  year = {2015},
  month = {Mar},
  publisher = {American Physical Society},
  doi = {10.1103/PhysRevLett.114.113002},
  url = {https://link.aps.org/doi/10.1103/PhysRevLett.114.113002}
}

@article{jepsen2020spin,
  title = {Spin transport in a tunable Heisenberg model realized with ultracold atoms},
  volume = {588},
  ISSN = {1476-4687},
  url = {http://dx.doi.org/10.1038/s41586-020-3033-y},
  DOI = {10.1038/s41586-020-3033-y},
  number = {7838},
  journal = {Nature},
  publisher = {Springer Science and Business Media LLC},
  author = {Jepsen,  Paul Niklas and Amato-Grill,  Jesse and Dimitrova,  Ivana and Ho,  Wen Wei and Demler,  Eugene and Ketterle,  Wolfgang},
  year = {2020},
  month = Dec,
  pages = {403–407}
}

@article{jepsen2021transverse,
  title = {Transverse Spin Dynamics in the Anisotropic Heisenberg Model Realized with Ultracold Atoms},
  author = {Jepsen, Paul Niklas and Ho, Wen Wei and Amato-Grill, Jesse and Dimitrova, Ivana and Demler, Eugene and Ketterle, Wolfgang},
  journal = {Phys. Rev. X},
  volume = {11},
  issue = {4},
  pages = {041054},
  numpages = {18},
  year = {2021},
  month = {Dec},
  publisher = {American Physical Society},
  doi = {10.1103/PhysRevX.11.041054},
  url = {https://link.aps.org/doi/10.1103/PhysRevX.11.041054}
}

@article{satzinger2021realizing,
  title = {Realizing topologically ordered states on a quantum processor},
  volume = {374},
  ISSN = {1095-9203},
  url = {http://dx.doi.org/10.1126/science.abi8378},
  DOI = {10.1126/science.abi8378},
  number = {6572},
  journal = {Science},
  publisher = {American Association for the Advancement of Science (AAAS)},
  author = {Satzinger,  K. J. and Liu,  Y.-J and Smith,  A. and Knapp,  C. and Newman,  M. and Jones,  C. and Chen,  Z. and Quintana,  C. and Mi,  X. and Dunsworth,  A. and others},
  year = {2021},
  month = Dec,
  pages = {1237–1241}
}

@article{haghshenas2026digital,
   title = {Digital quantum magnetism on a trapped-ion quantum computer},
  volume = {653},
  ISSN = {1476-4687},
  url = {http://dx.doi.org/10.1038/s41586-026-10445-3},
  DOI = {10.1038/s41586-026-10445-3},
  number = {8113},
  journal = {Nature},
  publisher = {Springer Science and Business Media LLC},
  author = {Haghshenas,  R. and Chertkov,  E. and Mills,  M. and Kadow,  W. and Lin,  S.-H. and Chen,  Y. H. and Cade,  C. and Niesen,  I. and Begušić,  T. and Rudolph,  M. S. and others},
  year = {2026},
  month = Apr,
  pages = {56–62} }

@article{bombin2012strong,
  title = {Strong Resilience of Topological Codes to Depolarization},
  author = {Bombin, H. and Andrist, Ruben S. and Ohzeki, Masayuki and Katzgraber, Helmut G. and Martin-Delgado, M. A.},
  journal = {Phys. Rev. X},
  volume = {2},
  issue = {2},
  pages = {021004},
  numpages = {10},
  year = {2012},
  month = {Apr},
  publisher = {American Physical Society},
  doi = {10.1103/PhysRevX.2.021004},
  url = {https://link.aps.org/doi/10.1103/PhysRevX.2.021004}
}

@article{valenti2019hamiltonian,
  title = {Hamiltonian learning for quantum error correction},
  author = {Valenti, Agnes and van Nieuwenburg, Evert and Huber, Sebastian and Greplova, Eliska},
  journal = {Phys. Rev. Res.},
  volume = {1},
  issue = {3},
  pages = {033092},
  numpages = {14},
  year = {2019},
  month = {Nov},
  publisher = {American Physical Society},
  doi = {10.1103/PhysRevResearch.1.033092},
  url = {https://link.aps.org/doi/10.1103/PhysRevResearch.1.033092}
}

@article{lewis1967classical,
  title = {Classical and Quantum Systems with Time-Dependent Harmonic-Oscillator-Type Hamiltonians},
  author = {Lewis, H. R.},
  journal = {Phys. Rev. Lett.},
  volume = {18},
  issue = {13},
  pages = {510--512},
  numpages = {0},
  year = {1967},
  month = {Mar},
  publisher = {American Physical Society},
  doi = {10.1103/PhysRevLett.18.510},
  url = {https://link.aps.org/doi/10.1103/PhysRevLett.18.510}
}

@article{gungordu2012dynamical,
  title = {Dynamical invariants for quantum control of four-level systems},
  author = {G\"ung\"ord\"u, Utkan and Wan, Yidun and Fasihi, Mohammad Ali and Nakahara, Mikio},
  journal = {Phys. Rev. A},
  volume = {86},
  issue = {6},
  pages = {062312},
  numpages = {9},
  year = {2012},
  month = {Dec},
  publisher = {American Physical Society},
  doi = {10.1103/PhysRevA.86.062312},
  url = {https://link.aps.org/doi/10.1103/PhysRevA.86.062312}
}

@article{truppe2017molecules,
  title={Molecules cooled below the Doppler limit},
  author={Truppe, S and Williams, HJ and Hambach, M and Caldwell, L and Fitch, NJ and Hinds, EA and Sauer, BE and Tarbutt, MR},
  journal={Nature Physics},
  volume={13},
  number={12},
  pages={1173--1176},
  year={2017},
  publisher={Nature Publishing Group UK London},
  url={https://doi.org/10.1038/nphys4241}
}

@article{ruttley2025long,
  title={Long-lived entanglement of molecules in magic-wavelength optical tweezers},
  author={Ruttley, Daniel K and Hepworth, Tom R and Guttridge, Alexander and Cornish, Simon L},
  journal={Nature},
  volume={637},
  number={8047},
  pages={827--832},
  year={2025},
  publisher={Nature Publishing Group UK London},
  url={https://doi.org/10.1038/s41586-024-08365-1}
}

@dataset{van_lomwel_data,
  author       = {van Lomwel, Alexander},
  title        = {Data for van Lomwel et al., Quantum simulation of
                   circular cluster interactions in a linear spin
                   chain
                  },
  month        = aug,
  year         = 2026,
  publisher    = {Zenodo},
  doi          = {10.5281/zenodo.22094112},
  url          = {https://doi.org/10.5281/zenodo.22094112},
}

\appendix

\newcommand{\PauliString}[4]{%
  \ensuremath{(\mathbf{#1}\mathbf{#2})_{#3,#4}}%
}

\setcounter{equation}{0}
\renewcommand{\theequation}{A\arabic{equation}}

\onecolumngrid
\begin{center}
    \large\bfseries End Matter
\end{center}
\twocolumngrid

\paragraph*{Lie algebra of the system.} Before evaluating the Lie algebra generated by the terms in the full system Hamiltonian (Eq.~\eqref{eq:systemHam}), it is first instructive to consider the case without boundary $X$ drives.  
The shorthand notation Eq.~\eqref{eq:stabilizer} is generalized to
\begin{equation}
    \stabb{A}{B}{j}{\alpha}= A_j\Bigl(\prod _{k=1}^{\alpha-1}Z_{j+k}\Bigr)B_{j+\alpha}\ ,
\end{equation}
with $j\in[1,n-\alpha]$, $\alpha\in[1,n-1]$, and the conventions $\stab{X}{j}{\alpha}\equiv \stabb{X}{X}{j}{\alpha}$ and $\stab{Y}{j}{\alpha}\equiv\stabb{Y}{Y}{j}{\alpha}$.
The system Hamiltonian, without the boundary $X$ driving, is comprised of the operator set
\begin{equation}
    \left\{\stab{X}{j}{1} + \stab{Y}{j}{1} \ , \ Z_i\right\} \ ,
    \label{eq:initialset}
\end{equation}
with $i\in[1,n]$.
The interactions $\stab{X}{j}{\alpha} + \stab{Y}{j}{\alpha}$ and $\stabb{X}{Y}{j}{\alpha}- \stabb{Y}{X}{j}{\alpha}$ are generated as Lie algebra elements by a systematic path of commutations between the terms in Eq.~\eqref{eq:initialset} and nested commutators of these terms. 
With the notation $D_j^\alpha=\stabb{X}{Y}{j}{\alpha}- \stabb{Y}{X}{j}{\alpha}$, these terms are generated recursively by alternating the commutators
\begin{equation}
    [\stab{X}{j}{\beta} + \stab{Y}{j}{\beta}, Z_j] \propto D_j^\beta \ ,
\end{equation}
for $\beta\in[1,\alpha]$, and
\begin{equation}
    [\stab{X}{j}{1} + \stab{Y}{j}{1},D_{j+1}^\beta] \propto \stab{X}{j}{\beta+1} + \stab{Y}{j}{\beta+1} \ ,
\end{equation}
for $\beta\in [1,\alpha-1]$, without the possibility of generating any other operator.
Thus, the full Lie algebra contains the elements
\begin{equation}
    \left\{\stab{X}{j}{\alpha} + \stab{Y}{j}{\alpha} \ , \ \stabb{X}{Y}{j}{\alpha}- \stabb{Y}{X}{j}{\alpha} \ , \  Z_i\right\} \ .
    \label{eq:liealgebrasusbet}
\end{equation}
With both $\stab{X}{j}{\alpha} + \stab{Y}{j}{\alpha}$ and $\stabb{X}{Y}{j}{\alpha}- \stabb{Y}{X}{j}{\alpha}$ comprising of $n(n-1)/2$ elements, and $n$ elements $Z_i$, there are a total of $n^2$ elements.
Importantly, this Lie algebra does not contain the isolated cluster interactions $\stab{X}{j}{\alpha}$, but instead the interactions $\stab{X}{j}{\alpha} + \stab{Y}{j}{\alpha}$ which conserve the total magnetization. 

With the boundary $X$ drives, the commutation between $X_1\equiv \stabb{Z}{X}{1}{0}$ and $Z_1$ yields the operator $Y_1\equiv \stabb{Z}{Y}{1}{0}$.
Successively longer boundary strings are generated recursively according to
\begin{align}
    &[\stab{X}{k}{1} + \stab{Y}{k}{1}, \stabb{Z}{X}{1}{k-1}] \propto \stabb{Z}{Y}{1}{k} \ , \label{eq:term1} \\
    &[\stab{X}{k}{1} + \stab{Y}{k}{1}, \stabb{Z}{Y}{1}{k-1}] \propto \stabb{Z}{X}{1}{k} \label{eq:term2} \ ,
\end{align}
for $k\in[1,n-1]$.
Commutations between Eq.~\eqref{eq:term1}, Eq.~\eqref{eq:term2} can then generate the terms $\stab{X}{k}{1}$,
\begin{equation}
    [\stabb{Z}{X}{1}{k}, \stabb{Z}{Y}{1}{k-1}] \propto \stab{X}{k}{1} \ .
\end{equation}
The same construction can be mirrored starting from $X_n$, generating the corresponding $\stab{X}{k}{1}$ terms from the right boundary and recursively extending them into the bulk, rather than from the left boundary as outlined above.
The inclusion of the boundary terms $X_1$, $X_n$ have thus effectively separated $\stab{X}{j}{1} + \stab{Y}{j}{1}$ into $\stab{X}{j}{1}$ and $\stab{Y}{j}{1}$ individually.
The Lie algebra generated by the operators that comprise the system Hamiltonian (Eq.~\eqref{eq:systemHam}) is thus entirely equivalent to the Lie algebra generated by the set of operators
\begin{equation}
    \left\{\stab{X}{k}{1} \ , \ Z_i \ , \ X_1 \ , \ X_n\right\} \ .
\end{equation}
This Lie algebra is the extended transverse field Ising Lie algebra \cite{orozco2024quantum,van2026fast} and has dimension $2n^2+3n+1$.
The set contains all the isolated OBC cluster interactions $\stab{X}{j}{\alpha}$ and $\stab{Y}{j}{\alpha}$, thus encompassing the target Hamiltonians $\HOBC$ (Eq.~\eqref{eq:generalizedcluster}) and $\HPBCxy$ (Eq.~\eqref{eq:XYtarget}), and notably the $Z$ parity operator $\mathbf{Z}$ included in the initial Hamiltonian $\Hinit$ (Eq.~\eqref{eq:Kinit}).
Refer to Ref.~\cite{van2026fast} for the details for how $\Hinit$ is derived for this specific Lie algebra.

\paragraph*{Infidelities and bounds.} Here, we specify the specific infidelity definitions used for the results in this Letter, along with a brief explanation of evaluability. 
Refer to Refs.~\cite{orozco2024quantum,van2026fast} for further details.
The infidelity between the evolved parent Hamiltonian $\Hevofinal$ and the target parent Hamiltonian $\Htarget$, $\mathcal{J}(\Hevofinal,\Htarget)$, is given by the function
\begin{equation}
    \mathcal{J}(A,B) = 1-\frac{\tr (AB)}{\sqrt{\tr (A^2)\tr (B^2)}} \ .
    \label{eq:opJ}
\end{equation}
Given that the system is comprised of a polynomially scaling Lie algebra, $\mathcal{J}(\Hevofinal,\Htarget)$ can be evaluated in terms of polynomially scaling coefficient vectors corresponding to the Lie algebra elements. 

For the preparation of the non-degenerate ground state $\ket{\Psi_\mathbf{T}}$ of the target Hamiltonian $\Htarget=\HPBCxy$ (Eq.~\eqref{eq:XYtarget}), the ground state infidelity $\mathcal{J}_\mathrm{gs}$ and the upper bound to this infidelity $\mathcal{B}_\mathrm{gs}$ are given by
\begin{align}
    \mathcal{J}_\text{gs}&= 1-\left|\braket{\Psi(T)}{\Psi_\mathbf{T}}\right|^2 
    \label{eq:gsJ} \ , \\
    \mathcal{B}_\text{gs}&= \frac{\bra{\Psi(T)}\Htarget \ket{\Psi(T)} +n}{2} 
    \label{eq:gsbound} \ ,
\end{align}
where $\ket{\Psi(T)}$ is the ground state of the evolved operator $\Hevofinal$.
Derived in~\cite{orozco2024quantum}, $\mathcal{B}_\text{gs}$ can be evaluated with polynomially scaling effort, via the backwards propagation of $\Htarget$ in a low-dimensional subspace owing to the polynomial scaling of the Lie algebra, whereas $\mathcal{J}_\text{gs}$ requires the full exponentially scaling statevectors.

For the preparation of the $2^\alpha$-fold degenerate ground space of the target Hamiltonian $\Htarget=\HOBC$ (Eq.~\eqref{eq:generalizedcluster}), the corresponding infidelity and infidelity bounds are given by
\begin{align}
    \mathcal{J}_\text{proj}&= \mathcal{J}(\mathsf{P}_\mathbf{T}, \mathsf{P}(T)) \label{eq:projJ} \ , \\
    \mathcal{B}_\text{proj}&=\frac{2^{-\alpha}\tr (\Htarget \mathsf{P}(T)) + n-\alpha}{2} \ ,
    \label{eq:projBound}
\end{align}
where $\mathsf{P}(T)$/$\mathsf{P}_\mathbf{T}$ is the projector onto  $2^\alpha$-dimensional ground space of $\Hevofinal$/$\Htarget$.
As with $\mathcal{B}_\text{gs}$ (Eq.~\eqref{eq:gsbound}), $\mathcal{B}_\text{proj}$ can be evaluated with polynomially scaling effort, whereas $\mathcal{J}_\text{proj}$ requires the full projectors.

\paragraph*{Periodic boundary conditions.} Here, we show that, by appropriately choosing the parameter $s=\pm1$, the ground state of the target Hamiltonian $\HPBCxy$ (Eq.~\eqref{eq:XYtarget}) is the same as the ground state of the true PBC Hamiltonian $\HPBC$ (Eq.~\eqref{eq:generalizedcluster}) for even $\alpha$, and the Hamiltonian $\HPBC + {\bf Z}$ for odd $\alpha$.
To show this, it is useful to use the relations,
\begin{equation}
\prod_{j=1}^n\stab{X}{j}{\alpha} = \begin{cases}
    \um & \text{if } \alpha \ \text{odd} \\
    (-1)^n\mathbf{Z}  & \text{if } \alpha \ \text{even} \ ,
\end{cases}
\label{eq:relation1}
\end{equation}
\begin{equation}
\prod_{j=1}^{n-\alpha}\stab{X}{j}{\alpha}  \prod_{j=1}^{\alpha}\stab{Y}{j}{n-\alpha} = \begin{cases}
    -\mathbf{Z} & \text{if } \alpha \ \text{odd} \\
    (-1)^n\mathbf{Z}  & \text{if } \alpha \ \text{even} \ ,
\end{cases}
\label{eq:relation2}
\end{equation}
and
\begin{equation}
    \stab{Y}{j}{n-\alpha} \mathbf{Z}= - \stab{X}{n-\alpha +j}{\alpha} \ ,
    \label{eq:relation3}
\end{equation}
for $j\in[1,\alpha]$.
A ground state $\ket{\Psi}$ of Hamiltonian $\HPBC$ necessarily satisfies the stabilizer relations $\stab{X}{j}{\alpha}\ket{\Psi} = \ket{\Psi}$, for $j\in [1,n]$.
The ground state $\ket{\Psi'}$ of $\HPBCxy$ satisfies $\stab{X}{j}{\alpha}\ket{\Psi'} = \ket{\Psi'}$ for $j\in [1,n-\alpha]$ and $\stab{Y}{j}{n-\alpha}\ket{\Psi'} = -s\ket{\Psi'}$ for $j\in[1,\alpha]$.
Starting with even $\alpha$, relation Eq.~\eqref{eq:relation1} reveals the ground state $Z$ parity, $\mathbf{Z}\ket{\Psi}=(-1)^n\ket{\Psi}$.
Using that $(-s)^\alpha=1$ for even $\alpha$, and relation  Eq.~\eqref{eq:relation2}, yields $\mathbf{Z}\ket{\Psi'}=(-1)^n\ket{\Psi'}$, and thus $\ket{\Psi'}$ has the same $Z$ parity as the true ground state $\ket{\Psi}$.
Using the parity and relation Eq.~\eqref{eq:relation3} gives the stabilizer expression $\stab{X}{j}{\alpha}\ket{\Psi'} = s(-1)^n\ket{\Psi'}$ for $j\in[n-\alpha+1,n]$.
Thus, for $\ket{\Psi'}$ to satisfy the same stabilizer relations as $\ket{\Psi}$, one chooses $s=+1$ for even $n$ and $s=-1$ for odd $n$ in $\HPBCxy$ (Eq.~\eqref{eq:XYtarget}).

For odd $\alpha$, the ground state $\ket{\Psi}$ of the Hamiltonian $\HPBC + \mathbf{Z}$ has the parity $\mathbf{Z}\ket{\Psi}=-\ket{\Psi}$.
Using relation Eq.~\eqref{eq:relation2}, the ground state $\ket{\Psi'}$ satisfies $\mathbf{Z}\ket{\Psi'}=s^\alpha\ket{\Psi'}$, thus one chooses $s=-1$ for the ground state to have the correct $Z$ parity.
From the relation Eq.~\eqref{eq:relation3}, the stabilizer expression reads $\stab{X}{j}{\alpha}\ket{\Psi'} = s^{\alpha+1}\ket{\Psi'}$ for $j\in[n-\alpha+1,n]$.
Given that $\alpha+1$ is even, $\ket{\Psi'}$ always satsifies the correct stabilizer relations, $\stab{X}{j}{\alpha}\ket{\Psi'} = \ket{\Psi'}$.

\end{document}